\documentclass[12 pt]{article}
\usepackage[utf8]{inputenc}
\usepackage{bm}
\usepackage{color}
\usepackage{ulem}
\usepackage{amsmath} 
\usepackage{caption}
\usepackage{amssymb}
\usepackage{algorithm}
\usepackage{algorithmic}
\usepackage{graphicx}
\usepackage{geometry}
\title{Simulation-based transition density approximation for the inference of SDE models}

\author{Xin Cai$^a$, Jingyu Yang$^b$, Zhibao Li$^b$, Hongqiao Wang$^b$, Miao Huang$^{b*}$ \\
{\it $^a$School of Statistics and Mathematics} \\
{\it Shanghai Lixin University of Accounting and Finance} \\
{\it Shanghai 995 Shangchuan Road, Shanghai 201209, People’s Republic of China}\\
[2mm]
{\it $^b$School of Mathematics and Statistics} \\
{\it Central South University} \\
{\it Changsha 410083, People’s Republic of China}\\
[2mm]
{$^*$ Corresponding author: Miao Huang}\\
{\it School of Mathematics and Statistics} \\
{\it Central South University} \\
{\it Changsha 410083, People’s Republic of China}\\
{\it E-mail: huangmiao@csu.edu.cn}}

\date{}

\begin{document}
\maketitle
\maketitle {\flushleft\large\bf Abstract}

Stochastic Differential Equations (SDEs) serve as a powerful modeling tool in various scientific domains, including systems science, engineering, and ecological science. While the specific form of SDEs is typically known for a given problem, certain model parameters remain unknown. 
Efficiently inferring these unknown parameters based on observations of the state in discrete time series represents a vital practical subject.
The challenge arises in nonlinear SDEs, where maximum likelihood estimation of parameters is generally unfeasible due to the absence of closed-form expressions for transition and stationary probability density functions of the states. 
In response to this limitation, we propose a novel two-step parameter inference mechanism. This approach involves a global-search phase followed by a local-refining procedure.
The global-search phase is dedicated to identifying the domain of high-value likelihood functions, while the local-refining procedure is specifically designed to enhance the surrogate likelihood within this localized domain.
Additionally, we present two simulation-based approximations for the transition density, aiming to efficiently or accurately approximate the likelihood function. Numerical examples illustrate the efficacy of our proposed methodology in achieving posterior parameter estimation.


\section{Introduction}
Complex mathematical model has a wide application in many areas such as medicine, machinery and finance.
While real-word phenomena are often affected by random factors and accurate mathematical models can hardly depict the affect of the disturbance. 
For example, in systems science, engineering science and ecological science, dynamical systems are often driven by inner noise or outer disturbance. That is, the motion is affected by the external forces and the input from the driver, which be modeled as stochastic processes.
For this problem, SDE offers an approach  for modeling these phenomena with stochastic disturbance.  
In these problems,
SDEs represent random variability in that they produce a different  solution trajectory each time solved.

With the wide application of SDEs in science and engineer, researchers have an increasing interest in estimation of parameters of SDEs. As far, a variety of algorithms have been proposed  to estimate the parameters of SDEs based on observations at discrete time\cite{caballero2021quantifying,craigmile2023statistical,Tsai2005Maxmum,picchini2014inference,nielsen2000parameter}.
Maximum likelihood estimation(MLE) in processes of Ornstein-Uhhenbeck type is proposed based on the inversion of the characteristic function and the use of the classical or fractional discrete fast Fourier transform\cite{BO2011588ML}. 
Martingale Estimating Functions(MEF) derive M-estimator from the approximate martingale estimating function\cite{Masayuki2008MEF}, Kalman filtering etc. 
Another kind of methods are  Bayesian methods that perform inference on the posterior distribution of parameters. 
The ML methods and Bayesian methods relies on the likelihood which is derived from transition probability that are usually unknown. 
In some simple cases researchers can compute the ML estimates (or MAP estimates) in closed form, and more generally computational techniques such as Maximum A Posteriori (MAP), Laplace approximations, Markov Chain Monte Carlo (MCMC), and various other Monte Carlo methods, can be employed. 

This study introduces an innovative approach for parameter inference in SDE models, combining global-search and local-refining techniques. The global-search, employing Bayesian optimization (BO), efficiently identifies the high-likelihood function domain. Subsequently, the surrogate is refined through the addition of new data, determined by an optimal experimental design criterion.
To address the critical transition density learning within the likelihood function for parameter inference in SDEs, we propose two simulation-based approximations. 
The first relies on accurate but time-intensive conditional density estimation (CDE), while the second, based on the independent approximation and kernel density estimation (IA-KDE) technique, prioritizes efficiency over accuracy.
Our two-step mechanism begins with exploring the global domain of high likelihood function values using the efficient IA-KDE method in the BO procedure to establish a coarse surrogate of the log-likelihood function. 
In the second step, we design new points based on the coarse surrogate model, creating a refined GP model by incorporating the new points and their corresponding accurate CDE approximation of the logarithm-likelihood function values.

The paper is organized as follows: In Section \ref{set:motivation}, we provide a comprehensive review of the formulation of stochastic differential equations, coupled with an introduction to the maximizing likelihood method.
Two simulation-based transition density approximations are presented in Section \ref{set:approximation}. 
In Section \ref{set:two_step}, we introduce the two-step mechanism for the parameter inference within BO technique and optimal design strategies and the detail numerical implement is shown in \ref{set:num_imp}.
To demonstrate the precision and efficiency of the proposed method, two model inference examples are examined in Section \ref{set:num_exam}. Finally, concluding remarks are provided in Section \ref{set:conclusion}.

\section{Motivation and Problem setup}
\label{set:motivation}
In some context of practical modeling, we might know the parametric form of the SDE, but the value of parameters are unknown. A typical setting is that we have an SDE with a vector of parameters $\boldsymbol{\theta}=(\theta_1,\dots,\theta_d)$\cite{sarkka2019solin}:
\begin{align}
\label{eq:general_sde}
\mathbf{dx}=\bm{f}(\mathbf{x},t;\boldsymbol{\theta})\mathbf{d}t+\boldsymbol{L}(\mathbf{x},t;\boldsymbol{\theta})\mathbf{d}\boldsymbol{\beta}, \,\,\,\mathbf{x}(t_0)=\mathbf{x}_0.
\end{align}
$\bm{f}$ is the drift functions with depend on the parameters $\bm{\theta}$ and the diffusion matrix of the Brownian motion $Q(\boldsymbol{\theta})$ might also depend on the parameters. 
Additionally, we have a set of observations of the SDE. For example, we might have a set of known values of the state $\bm{x}(\tau)$ at a certain finite number of time points $\bm{y} = [\bm{x}(\tau_0), \dots, \bm{x}(\tau_i),\dots], \tau_i\in [0, T-1]$, or only partial observations of the state, and these observations might also be corrupted by noise.

In the case that we observe a finite number of values of the SDE, say, $\mathbf{x}\left(\tau_0\right),  \ldots, \mathbf{x}\left(\tau_{T-1}\right)$, due to the Markov properties of SDEs, we can write down the likelihood of the observed values given the parameters as follows:
\begin{equation}
\label{eq:likelihood}
L(\bm{y}|\bm{\theta}) := p\left(\mathbf{x}\left(\tau_0\right), \ldots, \mathbf{x}\left(\tau_{T-1}\right) \mid \boldsymbol{\theta}\right)=\prod_{k=1}^{T-1} p\left(\mathbf{x}\left(\tau_{k}\right) \mid \mathbf{x}\left(\tau_{k-1}\right), \boldsymbol{\theta}\right),
\end{equation}
where $p\left(\mathbf{x}\left(\tau_{k}\right) \mid \mathbf{x}\left(\tau_{k-1}\right), \boldsymbol{\theta}\right),k=1\dots T-1$ are the transition densities of the SDE. 
One of most popular methods for SDE parameter estimation is the maximum likelihood (ML) method. 
In the ML method, we wish to maximize the likelihood,  or, equivalently, maximize the log-likelihood:
\begin{equation}
\label{eq:loglike}
\begin{aligned}
l(\boldsymbol{\theta}) & =\log p\left(\mathbf{x}\left(\tau_0\right), \ldots, \mathbf{x}\left(\tau_{T-1}\right) \mid \boldsymbol{\theta}\right) \\
& =\sum_{k=1}^{T-1} \log p\left(\mathbf{x}\left(\tau_{k}\right) \mid \mathbf{x}\left(\tau_{k-1}\right), \boldsymbol{\theta}\right) .
\end{aligned}
\end{equation}

As depicted in Equations \eqref{eq:likelihood} and \eqref{eq:loglike}, the transition density plays a crucial role in likelihood-based inference.
For intricate SDEs, analytic expression of $p\left(\mathbf{x}\left(\tau_{k}\right) \mid \mathbf{x}\left(\tau_{k-1}\right), \boldsymbol{\theta}\right)$ is not tractable and various methods have been developed for its approximation\cite{picchini2014inference,tronarp2018iterative,garcia2016iterated,archambeau2011approximate}.
In Section \ref{set:approximation}, we propose two simulation-based approximation methods to approximate the transition density with different accuracy and computational efficiency purposes. 

\section{Simulation-based transition density approximation}
\label{set:approximation}

It is well-established that the domain associated with the high-value likelihood function captures our attention in the context of the inference problem.
A viable approach to approximate $p(\bm{x}(\tau_{i+1})|\bm{x}(\tau_{i}), \bm{\theta})$, the important ingredient of likelihood, involves generating samples $\bm{x}(\tau_{i+1})$ through iterative execution of the discrete stochastic differential equation form, utilizing the specified model parameters $\bm{\theta}$ and initial condition $\bm{x}(\tau_{i})$.
Employing density estimation techniques\cite{chen2017tutorial,ambrogioni2017kernel,rothfuss2019conditional} becomes advantageous for the approximation of $p(\bm{x}(\tau_{i+1})|\bm{x}(\tau_{i}), \bm{\theta})$ in this context. 
Nevertheless, the likelihood maximization process necessitates numerous calls to the density estimation model, each with distinct model parameter candidates $\bm{\theta}$ and initial conditions $\bm{x}(\tau_i)$, resulting in a very large amount of computation.

Given a specific model parameter $\bm{\hat{\theta}}$ and initial condition $\bm{x}(\tau_{i}) = \bm{z}^0$, samples $\bm{x}(\tau_{i+1})= \bm{z}^1$ can be obtained by running the discrete SDE numerical form.
With $M$ repeated simulation, we can obtain the $M$ pair sample data $\mathcal{D} = \{\bm{z}^{0}_i, \bm{z}^{1}_i \}_{i=1,\dots,M}$, where $\bm{z}^{0}$ represents the value $\bm{x}(\tau_i)$ takes and $\bm{z}^{1}$ represents value $\bm{x}(\tau_{i+1})$ obtained.
It is worth emphasizing that repeated simulations can use computer parallelism to speed up processing.
The sample data exhibit characteristics corresponding to the model parameters $\bm{\theta}$, and a logical step is to derive an approximate transition density $p(\bm{z}^1|\bm{z}^0)$ from these data. 
In this context, we present two approximation approaches: one relies on known conditional density estimation (CDE) \cite{ambrogioni2017kernel}, and the other is based on independent approximation coupled with kernel density estimation, denoted as IA-KDE.

\subsection{Conditional Density Estimation (CDE)}
\label{set:cde}
Let $(\bm{z}^0, \bm{z}^1)$ be a pair of random variables. Let $p(\bm{z}^1 \mid \bm{z}^0)=p(\bm{z}^1, \bm{z}^0) / p(\bm{z}^0)$ denote the conditional probability density of $\bm{z}^1$ given $\bm{z}^0$. Typically, $Z^1$ is referred to as a dependent variable (explained variable) and $Z^0$ as conditional (explanatory) variable. Given a dataset of sample data $\mathcal{D}=\left\{\left(\bm{z}^0_i, \bm{z}^1_i\right)\right\}_{i=1}^N$ drawn from the joint distribution $\left(\bm{z}^0_i, \bm{z}^1_i\right) \sim p(\bm{z}^0, \bm{z}^1)$, the aim of conditional density estimation (CDE) is to find an estimate $\hat{f}_{NN}(\bm{z}^1 \mid \bm{z}^0)$ of the true conditional density $p(\bm{z}^1 \mid \bm{z}^0)$.

In the context of $\mathrm{CDE}$, the KL-divergence objective is expressed as expectation over $p(\bm{z}^0)$ :
\begin{equation}
\label{eq:cde_loss}
\mathbb{E}_{p(\bm{z}^0)}\left[\mathcal{D}_{K L}(p(\bm{z}^1 \mid \bm{z}^0) \| \hat{f}_{NN}(\bm{z}^1 \mid \bm{z}^0))\right]=\mathbb{E}_{p(\bm{z}^0, \bm{z}^1)}\left[\frac{\log p(\bm{z}^1 \mid \bm{z}^0)}{\log \hat{f}_{NN}(\bm{z}^1 \mid \bm{z}^0)}\right].
\end{equation}
Therefore, the optimal neural network-based conditional density model $\hat{f}_{NN}(\cdot \mid \cdot)$ is achieved by minimizing the expectation KL-divergence \eqref{eq:cde_loss}.
More details refer \cite{rothfuss2019noise,ambrogioni2017kernel}.

\subsection{Independent assumption for the conditional density approximation}
\label{set:sampling_based_kde}

Neural network-driven conditional density estimation techniques as mentioned in Section \ref{set:cde} demonstrate exceptional fitting capabilities, albeit requiring a substantial time investment for training. In this context, we introduce a sampling-based density estimation method characterized by rapid density assessment with a trade-off of lower accuracy. Despite the modest precision of the proposed approximation method, it retains the capacity to discern high and low likelihood values, thereby aiding in the identification of domains with high-value likelihood values.

We approximate the conditional density $p(\bm{z}^{1}|\bm{z}^{0}, \bm{\theta})$ by the product of approximate marginal density $q_I(\bm{z}^1|\bm{z}^0) = \prod_{k=1}^Dq(z^1_k|\bm{z}^0)$,
%$$q(z^1_k|\bm{z}^0) = \int p(\bm{z}_{\neq k}^1|\bm{z}^0, \bm{\theta})p(\bm{z}^0)d\bm{z}_{\neq k}^1,$$

$$q(z^1_k|\bm{z}^0) = \frac{1}{M}\sum_{i=1}^{M}K(z_k^1-z^1_{k,i}),$$
where $K(\cdot)$ is the kernel function and $\bm{z}^1_i$ are samples in $\mathcal{D}$ whose initial condition fall into the domain $[\bm{z}^0 - \bm{\epsilon}, \bm{z}^0 + \bm{\epsilon}]$, where $\bm{\epsilon}$ is fixed and preset threshold.
$\bm{z}_{\neq k}^1$ represents the variables without $z_k^1$.
Any kernel density estimation \cite{chen2017tutorial} technique can be easily used in the one-dimensional density approximation $q(z^1_k|\bm{z}^0)$.
\begin{equation}
\label{eq:sampling_de}
p(\bm{x}(\tau_{i+1})|\bm{x}(\tau_{i}), \bm{\theta})\approx q_I(\bm{x}(\tau_{i+1})|\bm{x}(\tau_{i}))  = \prod_{k=1}^Dq(x_k(\tau_{i+1})|\bm{x}(\tau_{i}))
\end{equation}

This sampling-based method own the advantages of efficiency and the drawbacks of low accuracy which effected by the dimensional independent approximate and the sampling-based density estimation which only can use insufficient samples.

\begin{figure}[H]
  \centering
	\includegraphics[width=.8\linewidth]{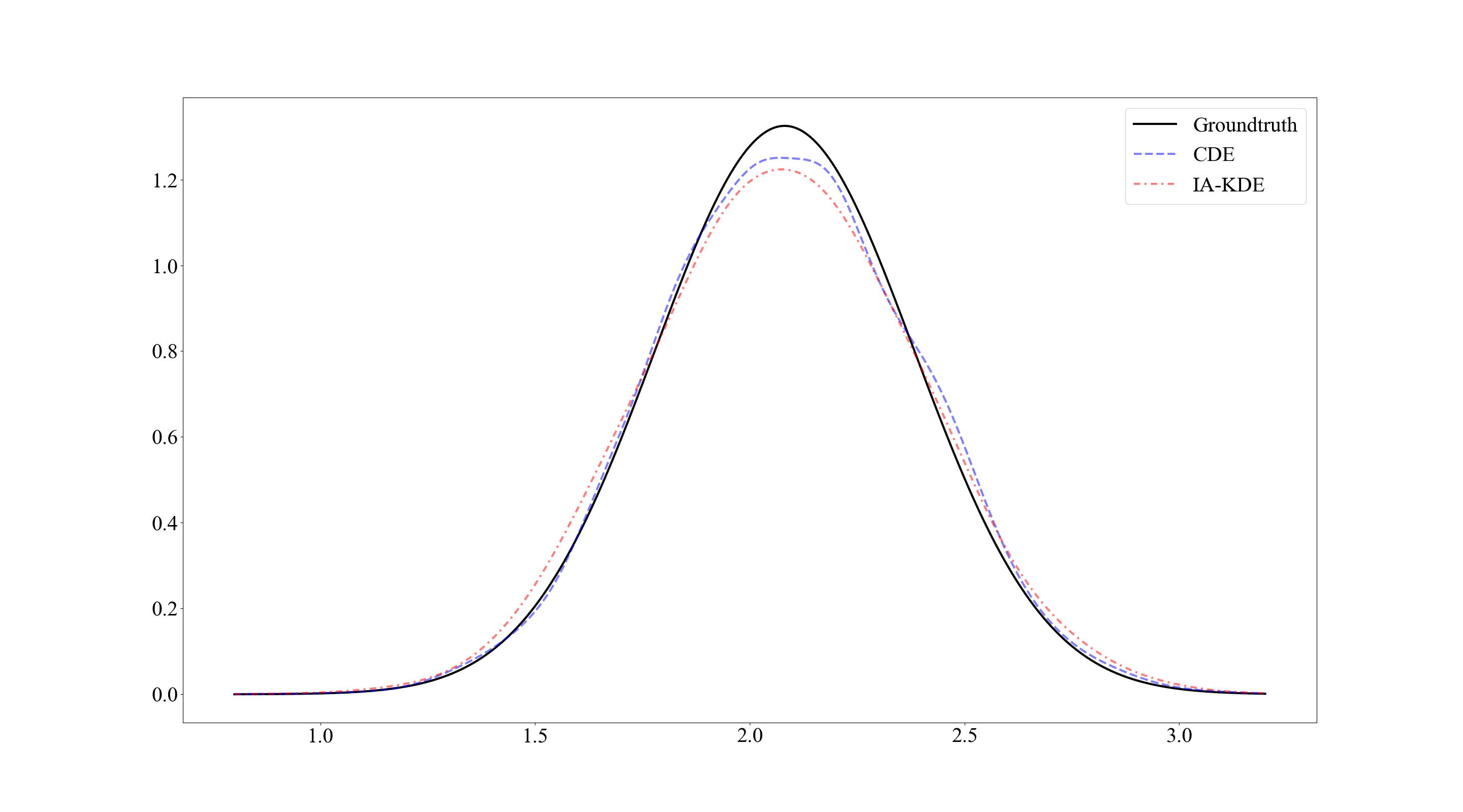}
	\caption{Comparison between CDE and IA-KDE in the Ornstein-Uhlenbeck process as discussed in Section.\ref{set:num_exam1}. The theoretical transition density is denoted by the black line, while the approximations computed by CDE and IA-KDS are represented by the red and blue lines, respectively.} 
	\label{fig:approximation_transition}
\end{figure}

While simulation-based CDE offers high accuracy, it comes with a time-costly drawback due to the neural network training process. 
On the other hand, IA-KDE, while less accurate, is highly efficient. 
The comparison about the  two approximations is shown in Figure \ref{fig:approximation_transition}.
The initial method relies on conditional density estimation, directly learning the transition density function from a substantial set of simulation data pairs denoted as $\{\bm{x}(\tau_i), \bm{x}(\tau_{i+1})\}_{1\le i \le T}$. This technique offers precise estimations owing to the robust fitting capabilities of neural networks; however, it necessitates a prolonged training period.
To enhance computational efficiency without compromising on accuracy, IA-KDE is a time-efficient density estimation method based on efficient sampling. 
Although this method may not achieve the highest precision, it leverages assumptions of dimensional independence and kernel density estimation for computational advantages over alternative techniques.
This proposed method proves valuable in Bayesian optimization, particularly in delineating the high-probability domain for maximizing likelihood estimation. 
Despite the inherent inaccuracy associated with maximum likelihood estimates, the valuable information they provide enables the exclusion of regions with low-value likelihood functions, thereby improving the ability to identify the global optimum.

%Given a specific model parameter $\bm{\hat{\theta}}$, samples $\bm{x}(\tau_{i+1})|\bm{x}(\tau_{i}), \bm{\theta}$ can be obtained by running the discrete SDE numerical form with the given $\bm{\hat{\theta}}$ with different  initial condition $\bm{x}(\tau_{i}) = \bm{z}_m, m=1,\dots,N_s$.
%With $m$ repeated simulation, we can obtain the $m\times N_s$ pair samples $\{\bm{z}^{0}_i, \bm{z}^{1}_i \}_{i=1,\dots,m\times N_s }$, where $\bm{z}^{0}$ represents $\bm{x}(\tau_i)$ and $\bm{z}^{1}$ represents $\bm{x}(\tau_{i+1})$.
%For obtaining the samples $x_k = \int p(\bm{x}_{\neq k}(\tau_{i+1})|\bm{x}(\tau_{i}), \bm{\theta})d\bm{x}_{\neq k}$, we can select the samples which is in the domain $[\bm{x}(\tau_{i}) - \bm{\epsilon}, \bm{x}(\tau_{i}) + \bm{\epsilon}]$, denoted as $\{\hat{\bm{z}}^{0}_i, \hat{\bm{z}}^{1}_i \}_{i=1,\dots}$, where $\bm{\epsilon}$
%is fixed and preset.
%A linear map can be learned efficiently $s$

\section{Efficient surrogate construction by two-step mechanism}
\label{set:two_step}
In real-world applications, our main emphasis is on analyzing the posterior distribution, denoted as $p(\bm{\theta}|\bm{y})$, rather than relying on a single point estimate, $\hat{\bm{\theta}}$. 
While the likelihood function is pivotal for inference, approximating it poses persistent challenges owing to inherent intractability or substantial computational expenses. 
Furthermore, optimizing complex likelihood functions becomes more intricate due to the existence of multiple local maxima.

In this context, we present a systematic two-step methodology for deducing posterior distributions. 
Initially, a thorough exploration of the domain is conducted, with a specific focus on areas characterized by a high-value likelihood function. 
Bayesian optimization (BO) employs an efficiently coarse approximation (IA-KDE) of the likelihood function to pinpoint the Maximum Likelihood (ML) estimation, denoted as $\bm{\theta}_{ML}$. 
This process concurrently allows for the encompassing of the domain associated with a high-value likelihood function. Subsequently, we enhance the surrogate likelihood function. To augment the Gaussian Process (GP) surrogate, we strategically introduce new design points and employ a precise simulation-based CDE method. This approach is designed to facilitate more accurate posterior inference.

\subsection{Bayesian optimization for maximum approximate likelihood estimation}

An approach to estimate the parameters $\boldsymbol{\theta}$ is to find the values that best fit the observed data. 
Due to the likelihood approximation in Section \ref{set:approximation} is time-costing and the $log$ likelihood function is continuous, we propose to find the maximum  likelihood value with respect to the parameter values by Bayesian optimization method. 

Bayesian optimization is an approach to globally optimizing objective functions that evaluation of objective function or the gradient is expensive. 
In this paper, we treat the log likelihood function $l: \bm{\theta} \to \mathbb{R}$ as a Gaussian process
\begin{equation}
l(\bm{\theta}) \sim \mathcal{GP}(\bm{0}, \Sigma(\bm{\theta},\bm{\theta}'))
\end{equation}

 Simply speaking, the function of interest $l(\bm{\theta})$ is a realization from a Gaussian random field, whose mean function is $m(\bm{\theta}),\bm{\theta}\in \mathbb{R}^d$ and covariance is specified by a kernel function $k(\bm{\theta},\bm{\theta}')$, namely,
 \begin{align}
     \text{Cov}[l(\bm{\theta}),l(\bm{\theta}')]=k(\bm{\theta},\bm{\theta}').
 \end{align}
The kernel $k(\bm{\theta},\bm{\theta}')$ is positive semidefinite and bounded.

Given data points $\{(\bm{\theta}_i,l_i)\}_{i=1}^n$, and one want to predict the value of $l$ at new point $\bm{\theta}$. We denote $\Theta:=[\bm{\theta}_1,\dots,\bm{\theta}_n]$, $\mathfrak{L}=[l_1,\dots,l_n]$, then the the joint distribution of $(\mathfrak{L},l^*)$ is Gaussian. That is,
\begin{align}
    \left[\begin{array}{c}
         \mathfrak{L}  \\
        l^* 
    \end{array}\right]
    \sim\mathcal{N}\left(
    \begin{array}{c}
        \bm{0}  \\
        0  
    \end{array},\left[\begin{array}{cc}
        K(\Theta,\Theta)+\sigma_n^2I & K(\Theta,\bm{\theta}^*) \\
        K(\bm{\theta}^*,\Theta) & K(\bm{\theta}^*,\bm{\theta}^*)
    \end{array}\right]\right),
\end{align}
where $\sigma_n^2$ is the variance of observation noise, $I$ is an identity matrix, and the notation $K(A,B)$ denotes the matrix of the covariance evaluated at all pairs of points in set $A$ and in set $B$ using the kernel function $k(\cdot,
\cdot)$. Then the conditional distribution $p(l^*|\bm{\theta}^*,\Theta,\mathfrak{L})$ is also Gaussian:
\begin{align}
    p(l^*|\bm{\theta}^*,\Theta,\mathfrak{L})=\mathcal{N}(\mu_{post},\Sigma_{post})
\end{align}
where the posterior mean and variance are:
\begin{align}
    \mu_{post}(\bm{\theta}^*)&=K(\bm{\theta}^*,\Theta)(K(\Theta,\Theta)+\sigma_n^2I)^{-1}\mathfrak{L})\\
    \Sigma_{post}(\bm{\theta}^*)&=K(\bm{\theta}^*,\bm{\theta}^*)-K(\bm{\theta}^*,\Theta)(K(\Theta,\Theta)+\sigma_n^2I)^{-1}K(\Theta,\bm{\theta}^*)
\end{align}

The covariance function determines the performance of GP in fitting data and prediction. We adopt the  RBF kernel here.
 
BO method comprises two main components: a probabilistic model: Gaussian process regression and an acquisition function. The process iterate the steps \cite{brochu2010tutorial}: (1) determine the point $\bm{\theta}_{n+1}\in\arg \max a_{p_{n}}(\bm{\theta})$; (2) evaluate the function value at $\bm{\theta}_{n+1}$; (3) update the Gaussian process regression surrogate with $\{(\bm{\theta}_1,l(\bm{\theta}_1))\dots,(\bm{\theta}_{n+1},l(\bm{\theta}_{n+1}))\}$. Typically, evaluations of the acquisition function $a$ are cheap compared to $l$ such that the optimization effort is negligible\cite{klein2017fast}.

The acquisition is to trade off exploration  vs. exploitation.   The acquisition function is the criterion to find future input candidate. Popular acquisition function includes Expected Improvement, Knowledge Gradient, Entropy Search and Predictive Entropy Search, etc..
Here we choose the EI (Expectated of improvement) function, which  performs robustly in most applications. It is defined as 
\begin{align}
    a_{\text{EI}}(\bm{\theta}|\mathcal{D}_n)=\mathbb{E}_p[\text{max}(l_{\text{min}}-l(\bm{\theta}),0)],
\end{align}
where $f_{\text{min}}$ is the best function value known. This function evaluate the expected drops over $f_{\text{min}}$. %经过一些简单的计算我们能发现，
the EI value is high where the mean is small or variance is large.

Utilizing efficient Bayesian optimization facilitates the acquisition of the maximum likelihood estimation $\hat{\bm{\theta}}_{ML}$ in conjunction with a surrogate model that demonstrates precision within the domain relative to the likelihood function of high value.

\subsection{Refine the GP surrogate}
\label{set:refine_GP}
Utilizing Bayesian optimization facilitates the acquisition of a domain characterized by a high-value likelihood function. The surrogate derived from the Bayesian optimization procedure exhibits reduced accuracy due to the coarse approximation employed. Achieving precise inference of the posterior distribution, denoted as $p(\bm{\theta}|\bm{y})$, necessitates the surrogates to demonstrate accuracy within the high probability domain. Consequently, enhancing the accuracy of the surrogate becomes imperative.

Utilizing the surrogate model acquired through Bayesian Optimization, we approximate a coarse log-likelihood function. Consequently, posterior samples can be generated using Markov Chain Monte Carlo (MCMC) methods based on this approximation. Once these samples are available, determining the support allows us to define $N_r$ points, denoted as $\bm{\theta}_1,\dots,\bm{\theta}_{N_r}$, through an optimal design criterion, such as Latin Hypercube Sampling (LHS) \cite{loh1996latin}.
In addition to LHS, random sampling from the obtained MCMC samples can also be treated as a design criterion which is used in \cite{wang2021inverse}. 
More sophisticated criteria rooted in information theory\cite{sebastiani2000maximum,foster2019variational} can be applied if deemed necessary for specific problems. 
Subsequently, simulations can be executed at these designated points, and the corresponding likelihood function values can be computed using the methods outlined in Section \ref{set:approximation}.

By incorporating the resulting data pairs into a refined Gaussian process surrogate, a new set of posterior samples can be drawn using MCMC methods. This global-search local-refined, two-step mechanism enhances the accuracy of the surrogate model and contributes to more reliable posterior inferences.

\subsection{Numerical implement}
\label{set:num_imp}
The posterior samples can be obtained by MCMC method from the product of approximate likelihood function and prior, 
\begin{equation}
\label{eq:post}
p(\bm{\theta}|\bm{y}) \propto \exp(l_{GP}(\bm{\theta}))p(\bm{\theta})
\end{equation}
Some known constraints about the model parameters, like positive, value range, etc,. can be specified in the prior $p(\bm{\theta})$.
Due to the GP surrogate is cheap, we can obtain the posterior samples easily by M-H sampler.

\begin{algorithm}[H]
    \caption{Global-search and local-refining parameter inference method}
    \label{alg:Two-step mechanism inference}
    \begin{algorithmic}[1]
    \STATE{\textbf{Inputs}:  Number of initial evaluation $N_0$, Number of maximum BO iteration$N_{max}$, Number of refined evaluation $N_r$, threshold value $\epsilon$;}
     \STATE{\textbf{Output}: The posterior samples $\{ \hat{\bm{\theta}}_i \}_{i=1,\dots,}$;}
     \STATE Randomly select $N_0$ points of $\{\bm{\theta}_i\}_{i=1,\dots,N_0}$ and compute the corresponding approximate logarithm likelihood value $\tilde{l}(\bm{\theta}_i)$ by IA-KDE in Section \ref{set:sampling_based_kde}.
 %\STATE Construct the initial Gaussian process regression and run the Bayesian optimization with the approximate likelihood function \eqref{eq:sampling_cde} until convergence. 
 \STATE  {\it // Global-search procedure}
\FOR {$N=1$ to $N_{max}$}	
	\STATE Compute/update the posterior probability distribution on $\tilde{l}$ using all available data 
     \STATE Let $\bm{\theta}_N$ be a maximizer of the acquisition function over $\bm{\theta}$, where the acquisition function is computed using
the current posterior distribution.
     \STATE Compute the value $\tilde{l}(\bm{\theta}_N)$ by IA-KDE approximation. 
     \STATE Increment $N$
     \IF{$|\bm{\bm{\theta}}_{N} - \bm{\bm{\theta}}_{N-1}|^2_2\le \epsilon$}
     	\STATE break;			     
     	\ENDIF
\ENDFOR
 \STATE  {\it // local-refine procedure}
\STATE Draw the posterior samples $\bm{\hat{\theta}}'$ by MCMC method based on the trained mean function of the GP surrogate.
\STATE Determine $N_r$ new design points based on the support of $\bm{\hat{\theta}}'$.
\STATE Take evaluation at the new design points and get the corresponding approximate log likelihood function values by CDE in Section. \ref{set:cde}.
\STATE Construct the refined surrogate by the $N_r$ new design points and its approximate log likelihood function values.
\STATE Draw the posterior samples $\{ \bm{\hat{\theta}}\}_{i=1,\dots,}$ by MCMC method based on the mean function of the refined GP surrogate.
   \end{algorithmic}
\end{algorithm}

Key points to note include:
\begin{enumerate}
\item Utilize $N_0$ initial data points, either randomly drawn from the prior distribution $p(\bm{\theta})$ or designed using LHS.
\item Employ the IA-KDE density estimation method in the Bayesian Optimization (BO) procedure for its efficiency. The final surrogate incorporates a neural network-based conditional density estimation.
\item $N_r$ new points can be generated concurrently through various optimal design strategies as discussed in Section \ref{set:refine_GP}. In this context, we adopt the widely applicable Latin Hypercube Sampling (LHS) and random sampling as our fundamental optimal design strategy.
\item Unless otherwise specified, we employ Euler–Maruyama method for the numerical simulation of SDEs in Equation \eqref{eq:general_sde}.
Detailed steps are as follows:
Give the initial condition $\hat{\bm{x}}(t_0)$ and the divide time interval $\left[t_0, t\right]$ into $M$ steps of length $\Delta t$. At each step $k$, do the following:\\
\begin{itemize}
\item[•] Draw random variable $\Delta \boldsymbol{\beta}_k$ from the distribution
$$
\Delta \boldsymbol{\beta}_k \sim \mathrm{N}(\mathbf{0}, \mathbf{Q} \Delta t) .
$$
\item[•] Compute
$$
\hat{\mathbf{x}}\left(t_{k+1}\right)=\hat{\mathbf{x}}\left(t_k\right)+\mathbf{f}\left(\hat{\mathbf{x}}\left(t_k\right), t_k\right) \Delta t+\mathbf{L}\left(\hat{\mathbf{x}}\left(t_k\right), t_k\right) \Delta \boldsymbol{\beta}_k .
$$
\end{itemize}
\end{enumerate}

\section{Numerical examples}
\label{set:num_exam}

\subsection{Ornstein-Uhlenbeck process}
\label{set:num_exam1}

\begin{figure}[H]
	\center
	\includegraphics*[width=12cm]{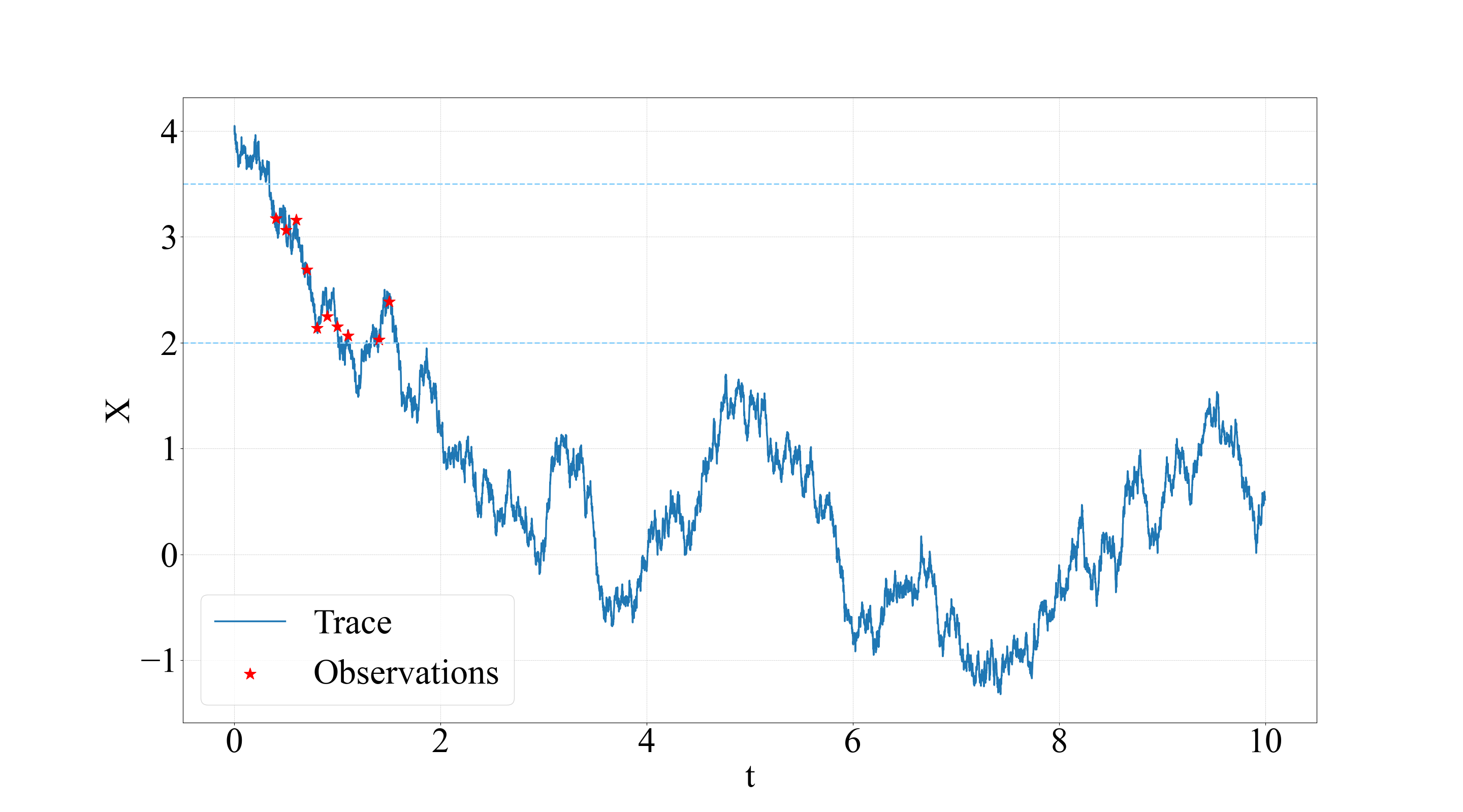}
	\centering
	\caption{Trajectory and observations in one simulation with $\lambda=1$. The blue trajectory are the solution computed by Euler-Maruyama scheme and the red stars are the observations used for inference.}\label{fig:exam1_obs}
\end{figure}

We first consider the Ornstein-Uhlenbeck process :
\begin{equation}
\label{eq:exam1_OU}
\mathrm{d}x=-\lambda x\mathrm{d}t+\mathrm{d}\beta,  
\end{equation}
with the unknown parameter $\lambda$ and $\mathrm{d}\beta$ is Brownian motion.
In numerical simulation, we set $T=10$, $N_{\Delta}=10000$, $\Delta t =T/N_{\Delta}=0.001$s. 
The observations are the state function value at $\tau_k$ time point with the truth parameter $\lambda= 1$. 
In this context, the time points $\tau_k, k=1,\dots,N$ represent a set of one hundred evenly distributed time points spanning the interval $[0, 10]$ along a SDE trajectory. 
For efficient inference, parts of observations which fall in the interval $[2, 3.5]$ are used in this example as shown in Figure \ref{fig:exam1_obs}.
$M=5\times 10^{3}$ pair of simulation samples are evaluated for the approximation of transition density.

\begin{figure}[H]
	\center
	\includegraphics*[width=12cm]{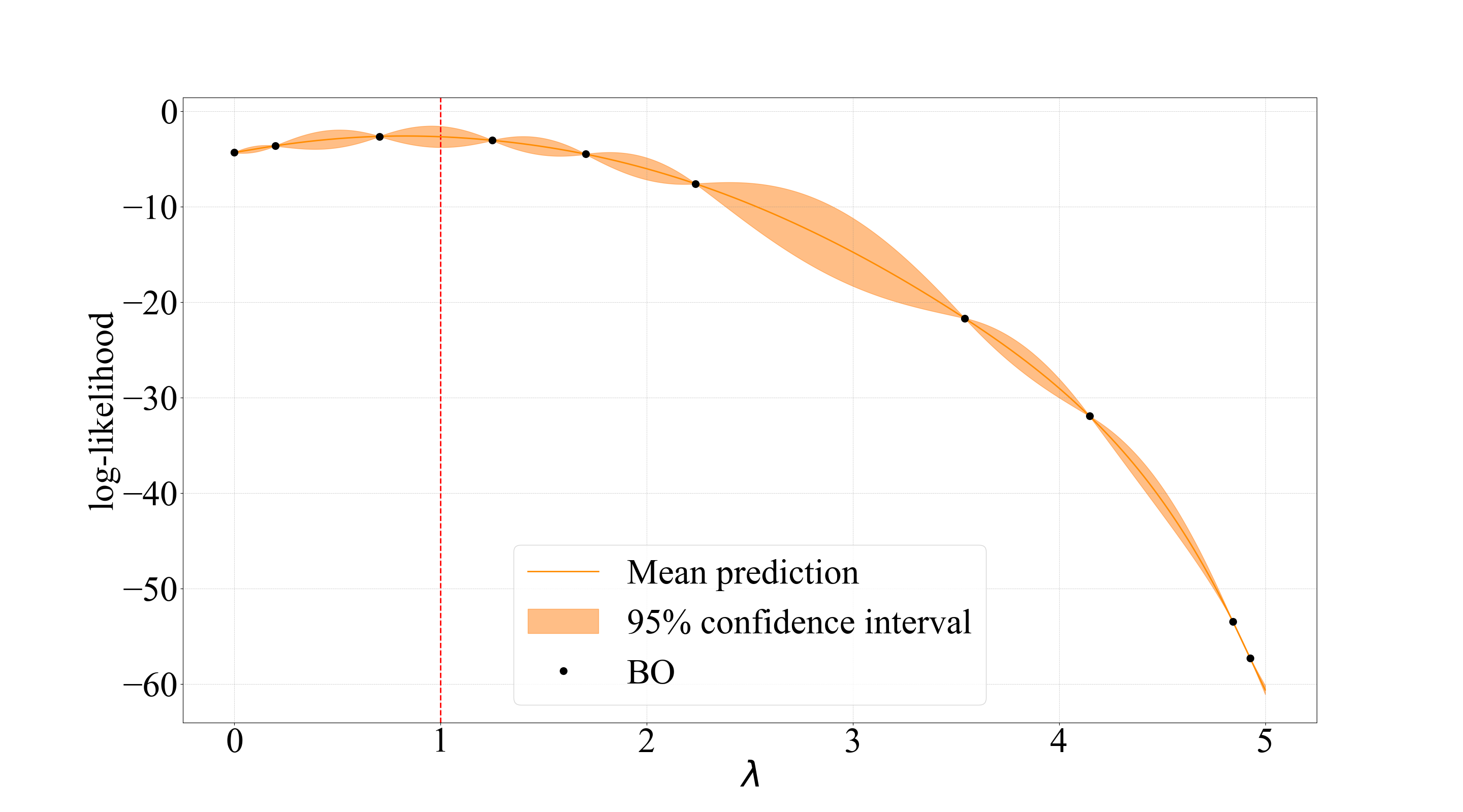}
	\centering
	\caption{The coarse approximation (GP surrogate) of log-likelihood function obtained in the BO procedure.}\label{fig:exam1_surrogate}
\end{figure}

In the BO procedure, a coarse GP surrogate, as illustrated in Figure \ref{fig:exam1_surrogate}, is obtained through the utilization of the IA-KDE approximation method detailed in Section \ref{set:sampling_based_kde}. The BO procedure incorporates five initial design points and five optimal design points. Analysis of Figure \ref{fig:exam1_surrogate} reveals that the domain of high-value log-likelihood functions is situated within the interval $[0.28, 1.62]$, representing the $90\%$ confidence interval of coarse posterior samples.

\begin{figure}[H]
	\center
	\includegraphics*[width=12cm]{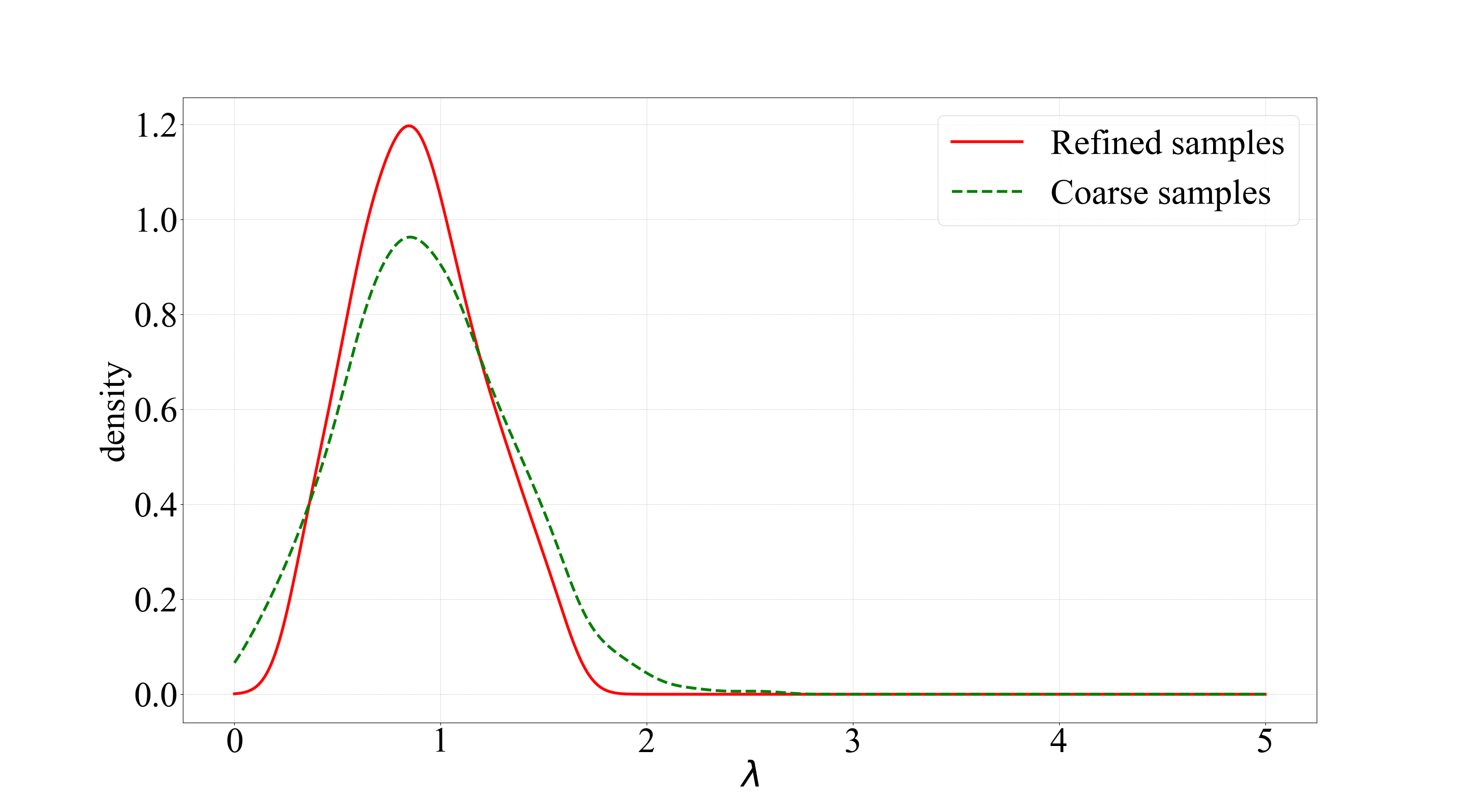}
	\centering
	\caption{The density estimation of the posterior samples computed by coarse GP surrogate (green dot) and the refined one (red line).}\label{fig:exam1_poste}
\end{figure}

Five refined selected design points are determined through LHS within the specified support interval of $[0.28, 1.62]$. Employing the CDE approximation during the refined phase enables the generation of an enhanced surrogate for the log-likelihood, along with its corresponding posterior samples. As depicted in Figure \ref{fig:exam1_poste}, the posterior estimation derived from both the coarse surrogate and the refined surrogate is presented. Notably, the refined surrogate demonstrates a substantial reduction in posterior variance when compared to the coarse approximation. The resultant posterior samples yield a mean value of $0.89$, closely approximating the ground truth of $1$, with a posterior variance of $0.09$.

\subsection{double-well model(1)}
\begin{figure}[H]
	\center
	\includegraphics*[width=12cm]{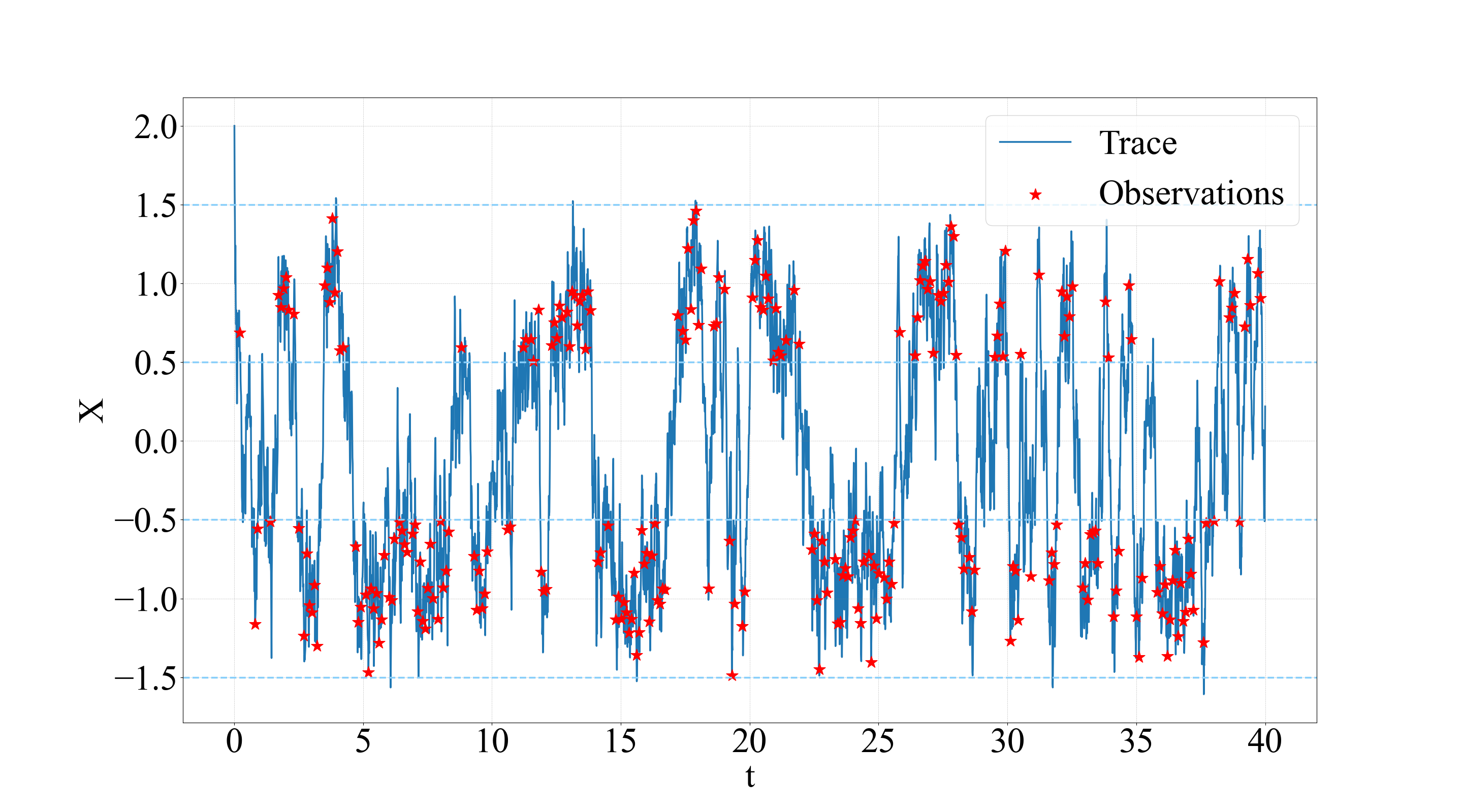}
%	\\
%	\includegraphics*[width=8cm]{Figure/exam2_sde2.jpg}
%	\includegraphics*[width=8cm]{Figure/exam2_sde3.jpg}
	\centering
	\caption{Trajectory and observations in one simulation with $\bm{\theta}=[3, -6]$. The blue trajectory are the solution computed by Euler-Maruyama scheme and the red stars are the observations used for inference.}\label{fig:exam2_sde}
\end{figure}

Consider
the following double-well model
\begin{align}
    \mathrm{d}x=(\theta_1 x+\theta_2 x^2)\mathrm{d}t+\sqrt{\max(4-1.25x^2,0)}\mathrm{d}\beta
\end{align}
where is a Brownian motion and $[\theta_1, \theta_2]$ are the unknown parameters. 
We employ the Euler–Maruyama scheme to generate a simulated dataset consisting of $M=4000$ samples, with a time step of $\Delta t=0.01$. Subsequently, we selectively retain every $10$th sample point as an observation, resulting in a total of $n=400$ observations with the truth parameter $\bm{\theta}=[3, -6]$. To enhance the efficiency of the inference process, we exclusively consider observations within the intervals $[-1.5, -0.5]$ and $[0.5, 1.5]$, as illustrated in Figure \ref{fig:exam2_sde}. Our primary goal is to deduce the values of the two parameters, denoted as $\bm{\theta}=[\theta_1,\theta_2]$, associated with the drift function.

\begin{figure}[H]
	\center
	\includegraphics*[width=10cm]{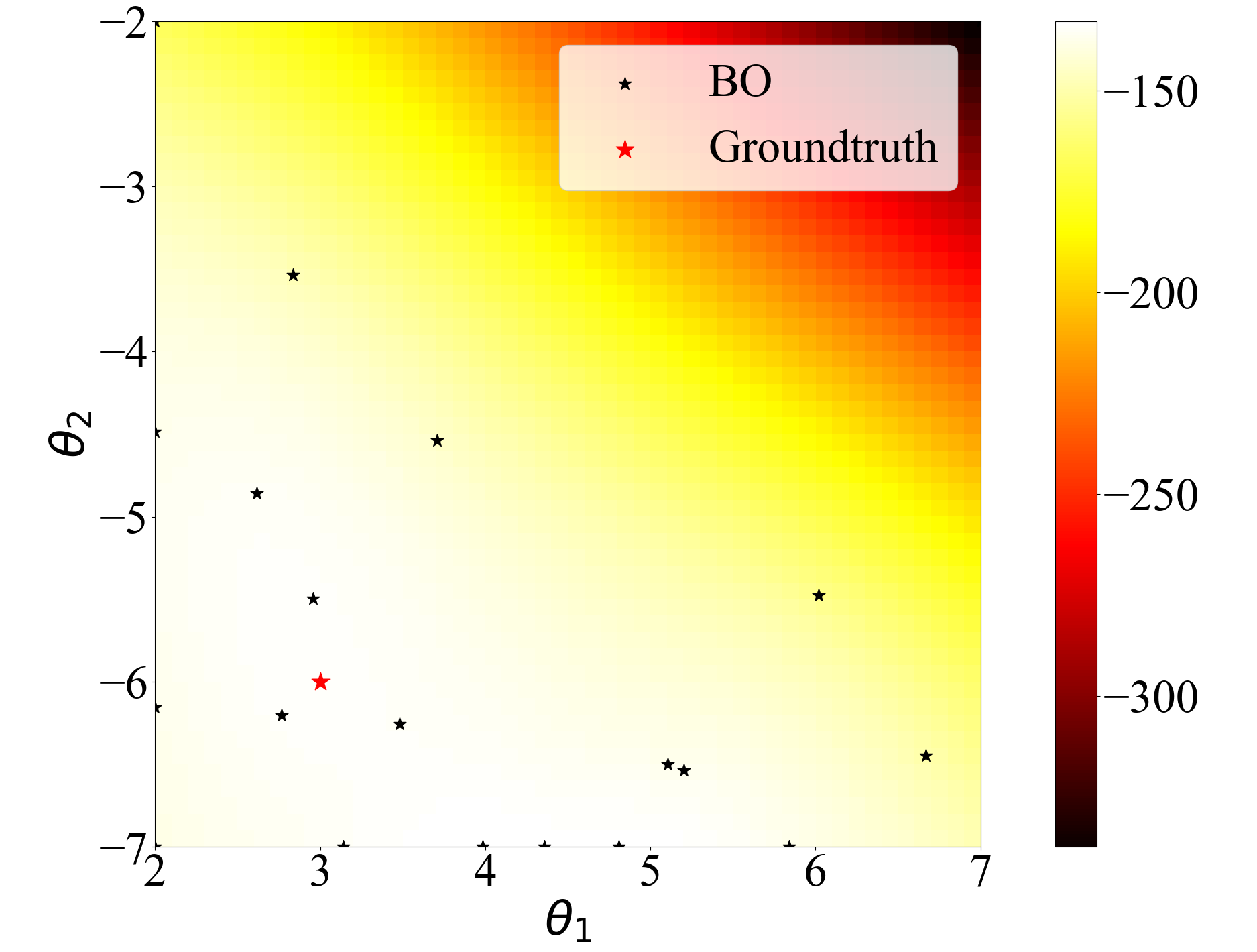}
	\centering
	\caption{The coarse approximation (GP surrogate) of log-likelihood function obtained in the BO procedure.}\label{fig:exam2_surrogate}
\end{figure}

\begin{figure}[H]
	\center
	\includegraphics*[width=10cm]{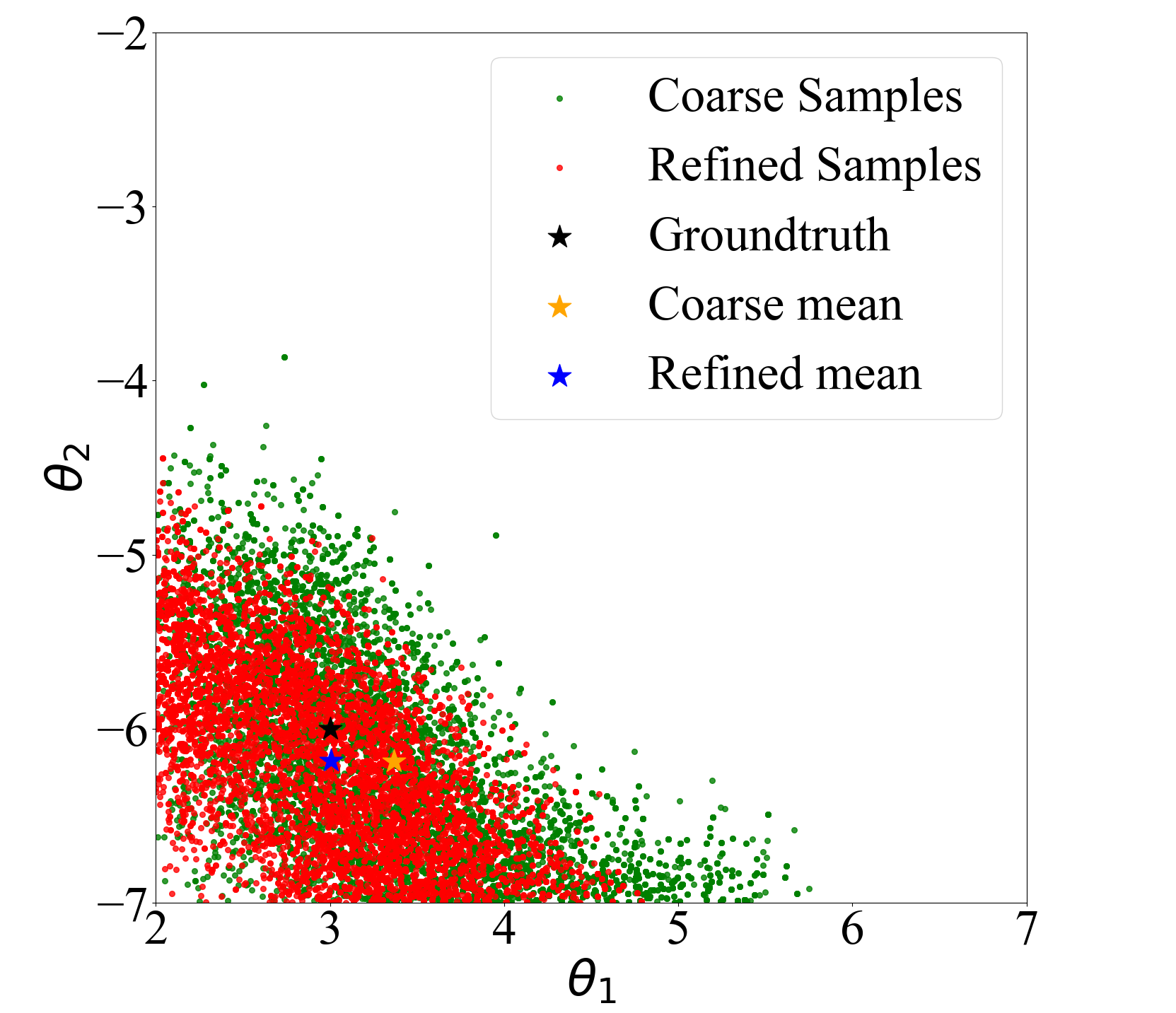}
	\centering
	\caption{Posterior samples drawn from the coarse GP surrogate (green dot) and the refined one (red dot).}\label{fig:exam2_postsamples}
\end{figure}

The design points are determined using the BO procedure, as depicted in Figure \ref{fig:exam2_surrogate}. Additionally, the posterior mean function is presented in the same figure, revealing that a majority of design points are situated within the high-value likelihood domain. Figure \ref{fig:exam2_postsamples} displays the posterior samples of both the coarse surrogate and the refined surrogate. Notably, the refined samples exhibit a posterior mean of $[2.97, -6.16]$, which is in closer proximity to the groundtruth $[3, -6]$ than the coarse samples' mean of $[3.37, -6.16]$. Furthermore, the variance of the refined samples, represented as $[0.35, 0.28]$, is smaller than that of the coarse samples, denoted as $[0.54, 0.43]$. This discrepancy underscores the effectiveness of the CDE approximation in correcting the posterior distribution.
%\begin{figure}[H]
%	\center
%	\includegraphics*[width=12cm]{Figure/exam2_NNCDE.jpg}
%	\centering
%	\caption{1}\label{g1-1}
%\end{figure}
%\begin{figure}[H]
%	\center
%	\includegraphics*[width=12cm]{Figure/exam2_particalCDE.jpg}
%	\centering
%	\caption{1}\label{g1-1}
%\end{figure}

\section{Conclusion}
\label{set:conclusion}

We present a two-step mechanism for the inference of SDE models with employing the approximations of transition density.
The high-value likelihood function domain is detected by the global-search phase and the GP surrogate is refined by the local-refine procedure.
Two simulation-based conditional density estimations are developed here for efficient or accurate approximation of the likelihood function.
To demonstrate the efficacy of our approach, we present two numerical examples that showcase its outstanding performance and reliability in different SDE models.

The incorporation of both the coarse Gaussian Process (GP) model and its refined counterpart is anticipated to enhance the efficiency and accuracy of surrogate construction. This strategic combination will undergo further investigation, as it is deemed a logical selection for optimization.

\section*{Acknowledgement}
Hongqiao Wang acknowledges the support of NSFC 12101615 and the Natural Science Foundation of Hunan Province, China, under Grant 2022JJ40567.
Zhibao Li acknowledges the support of NSFC 12271526.
This work was carried out in part using computing resources at the High Performance Computing Center of Central South University.

\bibliographystyle{plain}
\bibliography{ref}

\end{document}